\documentstyle[aps,prc,12pt,epsf]{revtex}

\newcommand{\be}{\begin{equation}}
\newcommand{\ee}{\end{equation}}
\newcommand{\ba}{\begin{eqnarray}}
\newcommand{\ea}{\end{eqnarray}}

\tightenlines
\begin{document}

\title{
Equation of State, Spectra and Composition of Hot and Dense Infinite Hadronic 
Matter in a Microscopic Transport Model\footnote{supported by BMBF, GSI, DFG 
and Graduiertenkolleg 'Schwerionenphysik'}
}

\author{
M. Belkacem$^1$\footnote{Alexander von Humboldt Fellow}, 
M. Brandstetter$^1$, S. A. Bass$^2$\footnote{Feodor Lynen Fellow of the 
Alexander von Humboldt Foundation},  
M. Bleicher$^1$\footnote{supported by the Josef Buchmann Foundation},
L. Bravina$^{1\dagger}$, \\
M. I. Gorenstein$^3$, J. Konopka$^1$, L. Neise$^1$,
C. Spieles$^1$, 
S. Soff$^1$, H. Weber$^1$, \\
H. St\"ocker$^1$  and W. Greiner$^1$
}

\address{
$^1$ Institut f\"{u}r Theoretische Physik, J. W. Goethe-Universit\"{a}t, \\
D-60054 Frankfurt am Main, Germany \\
$^2$ Dept. of Physics, Duke University, Durham 27708-0305, USA \\
$^3$ Bogolyubov Institute for Theoretical Physics, Kiev, Ukraine
}

\date{ \today }
\maketitle

\begin{abstract}

Equilibrium properties of 
infinite relativistic hadron matter are investigated using the
Ultrarelativistic Quantum Molecular Dynamics (UrQMD) model. The simulations are
performed in a box with periodic boundary conditions. Equilibration times
depend critically on energy and baryon densities.
Energy spectra of various hadronic species are shown to be
isotropic and consistent with a single temperature in equilibrium. The
variation of energy density versus temperature shows a Hagedorn-like behavior
with a limiting temperature of 130$\pm$10 MeV.
Comparison of abundances of different particle species to ideal hadron gas model 
predictions show good agreement only if detailed balance is implemented for all 
channels.
At low energy densities, high mass resonances 
are not relevant; however, their importance raises with increasing energy
density.
The relevance of these different conceptual
frameworks for any interpretation of experimental data is questioned.
\end{abstract}


\newpage


The intriguing possibilities of phase transitions at various orders in the 
nuclear equation of state (EoS) have long played a central role in heavy-ion 
physics from intermediate to very high energies \cite{sto-grei1,jaq,good}.
Indications about the nuclear liquid-gas phase transition at moderate
temperatures, $T < 10$ MeV, and densities, $\rho < \rho_0$, 
\cite{eos,aladin,multics,belkacem,peilert}, and about the 
transition to resonance matter at higher densities
and/or temperatures ($\rho \approx 3-5 \rho_0$, $T \approx 100-150$ MeV)
\cite{chapline,stoecker,boguta} have been reported.
Deconfinement of hadronic matter into quark-gluon plasma and strange matter 
may occur at even higher energy densities. Deconfinement temperatures of about 
$T_c \approx 150 \pm 10$ MeV have been predicted by lattice QCD simulations 
for zero net baryon - and strangeness density \cite{latqcd}.
However, the statistical concept of phase transitions is developed for 
stationary states close to equilibrium \cite{landau,balescu,huang,greiner}. 
Therefore, the question immediately 
arises:
can thermal and chemical equilibrium concepts actually be used and equilibrium
be reached in central heavy-ion
collisions?

At bombarding energies $E_{{\rm Lab}} \le$ 1 GeV/nucleon, experimental data 
on
multifragmentation \cite{multics1} have been described e.g. 
by statistical models
(Quantum Statistical Model (QSM) \cite{horst,konopka,gulminelli} or 
Statistical Multifragmentation Models (SMM) \cite{gross}). 
Also measured particle abundances and energy spectra from central heavy-ion
reactions at both AGS (14.6 GeV/nucleon) and SPS (200 GeV/nucleon) energies 
\cite{qm96} have been described
in statistical models assuming global thermal and hadrochemical equilibrium
(with overall moderate longitudinal and transverse expansions)
\cite{stat,stat2,stat1,goren}.
These models assume an instantaneous global freeze-out for all
particles. 

However, this hypothesis is - on first sight - 
not consistent with the results of
microscopic models. They predict different freeze-out times and radii 
for the different particle
species \cite{bravina,bass,urqmd,sorge,soff}. 
An other related question is, whether
the short time scales (10$^{-23}$s) are long enough for the system to move
through consecutive equilibrium states before freeze-out.

In this paper, we study equilibrium properties of infinite
relativistic hadronic matter within the framework of the Ultrarelativistic
Quantum Molecular Dynamics (UrQMD) model \cite{urqmd}. However, in contrast to
the usual simulations of heavy-ion collisions from the two colliding nuclei to
the complex final state, here a closed microcanonical system is constructed at 
a fixed net baryon density and net strangeness density, confined to a cubic box with 
periodic boundary conditions.
The yield of the various particle species can change due to inelastic
collisions and decays of hadron resonances and strings.

UrQMD, as a dynamical microscopic model, follows the time
evolution of a non-equilibrium A-particle system, e.g. a heavy-ion reaction, 
in the entire many-body phase space. It has
been applied in the
energy range from 100 MeV/nucleon up to several hundreds of GeV/nucleon for
heavy ion collisions 
using the same basic concepts and physics inputs at all energies. The model 
includes explicitly 55 different baryon species
(nucleons, deltas, hyperons, and their known resonances \cite{pdb} 
up to masses of 2.25 GeV) and 32 different meson species 
(including the known meson resonances \cite{pdb} up to masses
of 2 GeV), as well as their respective anti-particles and all
isospin-projected states \cite{urqmd}. For higher mass excitations, a 
string mechanism is invoked. Hadrons produced through string decays
have non-zero formation time $\tau_f$, 
which depends on energy-momentum of the particle.
Newly formed particles cannot interact during their formation time.
The leading hadrons with constituent quarks interact within their
formation time with a reduced cross section, which is taken to be 
proportional to the number of their
original constituent quarks.
All hadrons are propagated (in a relativistic cascade
sense) according to Hamilton's equations of motion, supplemented by a
relativistic Boltzmann-Uehling-Uhlenbeck collision
term involving all hadron states. The collision term is based on tabulated or 
parameterized experimental cross section (when available). Resonance 
absorption and scattering is handled via the 
principle of detailed balance. If no experimental information is
available, the cross section is either  calculated via
an OBE model or via a modified additive quark model.
The baryon-antibaryon annihilation cross section is parameterized as
the proton-antiproton annihilation cross section at the same center of mass 
energy.
For a detailed description of the UrQMD model, the reader is referred to the
original publication of the model presentation \cite{urqmd}.

For the present study, additional features have been added to the model, which
allow for the calculation of infinite hadronic matter properties in 
equilibrium. This is done by confining 
the
different particles in a cubic box with cyclic boundary conditions at a fixed
net baryon density $\rho_B$ and fixed net strangeness $\rho_S$. In this paper, 
we restrict our study to a
net baryon density 
$\rho_B = \rho_0 = 0.16$ fm$^{-3}$ and $\rho_S =$ 0 fm$^{-3}$, i.e. net 
strangeness zero. 
Although in heavy-ion collisions, $\rho_S \neq$ 0 and $\rho_B \neq \rho_0$ are
expected, we shift the detailed study of different net baryon
densities and/or non zero net strangeness to a forthcoming 
paper \cite{nxtpap}. 

The initial system considered here consists of 80 protons + 80 neutrons, 
which are uniformly distributed in configuration space in a
cubic box of $10\times10\times10$ fm$^3$. The momenta are uniformly
distributed in a sphere with random radius and then rescaled to
the desired energy density. 

Figure 1 displays the time evolution of the multiplicities (averaged over 50
events) of nucleons,
pions and kaons at two different energy
densities, $\epsilon = 200$ MeV/fm$^3$ (upper panel) and 
$\epsilon = 700$ MeV/fm$^3$ (lower panel). Particle
multiplicities saturate after some time. This equilibration time depends
strongly on energy density $\epsilon$. The different hadrons continue to 
interact strongly, but their average absolute numbers remain nearly constant. 
Note that the equilibration time
of kaons is much longer at the energy density $\epsilon =$ 0.2 GeV/fm$^3$
than those for nucleons and
pions, which are rather similar. In fact, this is the
case for all strange hadrons (both mesons and baryons).
This difference between the strange and non strange hadrons 
decreases with growing energy density. Equilibration
times of all hadrons are quite similar for energy densities above 0.7 
GeV/fm$^3$. 

We have checked whether our system 
is ergodic in the equilibrium phase: Both ensemble averages 
(i.e. when the averaging is done over a large set of parallel
events) and time averages (when the averaging is done over time evolution of a 
single event, but in both cases after equilibrium is established)
coincide within 1\%. The following results are obtained by time averaging 
(if not stated otherwise). All
quantities presented below have been extracted from the equilibrium phase, i.e.
after particle multiplicities have achieved their saturation values\footnote{In
practice, we start the time averaging much after particle multiplicities of the
most relevant species (N, $\pi$, $\Delta$, K, $\Lambda$) 
have achieved their saturation values, to allow for the
equilibration of the higher resonances. E.g., at $\epsilon =$ 0.2 GeV/fm$^3$,
the averaging starts after a time evolution of 500 fm/c, much after the kaons
have equilibrated ($\sim$ 250 fm/c, see Fig. 1). 
For energy densities larger than 0.5
GeV/fm$^3$, we average after an initial evolution of 150 fm/c. Moreover, 
in most cases,
the time averaging is done for a time evolution from 500 to 1000 fm/c.}
Figure 2 shows the energy spectra for four particle species; 
nucleons, pions, lambdas
and kaons (top panels) for two different energy
densities $\epsilon = 250$ MeV/fm$^3$ (left) and $\epsilon = 800$ 
MeV/fm$^3$ (right). The
distributions of the three momentum components $p_x$, $p_y$ and $p_z$ are given 
for nucleons in the bottom panels. The energy spectra (top panels) are 
reproduced by
Boltzmann distributions, ${\mathrm exp}(-E/T)$, with nearly the same 
temperature for all
4 species ($T=125\pm10$ MeV for $\epsilon=250$ MeV/fm$^3$, and $T=130\pm10$ MeV 
for $\epsilon=800$ MeV/fm$^3$). 

The momentum components distributions (lower part of the figure)
are reproduced by Gaussian distributions, ${\mathrm exp}(-p^2/2mT)$, with 
the same
temperature for the three space directions (complete isotropy of
the momentum distributions). This holds for all other hadron
species as well, if analogous procedures are applied. 
At this point, it should be noted that the temperature of mesons is 
always smaller (by 5 to 10\%) than that of the baryons. This is
mainly due to decay kinematics of resonances \cite{brandstetter}.
The temperature used in Fig. 2 is the
average value between the two.
Apart from this small discrepancy,
all particle have the same temperature $T$ and are globally equilibrated.  

Figure 3 shows the energy density
versus the temperature as resulting from the box calculations using the UrQMD 
model (open diamonds). It shows a
Hagedorn-like equation of state \cite{hage}: a rapid rise
of the temperature at low energy densities is followed by a saturation
at a temperature around 130$\pm$10 MeV. The calculations have 
been done up to energy densities of 5 GeV/fm$^3$, where still the same 
limiting temperature is observed. In the same figure, solid circles show the
EoS of the UrQMD model without strings and many-body decays. In this case, the
temperature shows a continuous rise with energy density and no limiting
temperature is observed. The inclusion of strings in the UrQMD model changes 
then the equation of
state from a continuously increasing temperature with energy
density (when strings are not taken into account) to a Hagedorn-like EoS with
a limiting temperature, $T_H = 130\pm10$ MeV.

Ideal hadron gas models have been used to describe particle abundances
and energy spectra from central heavy-ion reactions at both AGS and SPS
energies \cite{stat,stat2,stat1,goren}. 
In these models, the system 
is described
by a grand canonical ensemble of non-interacting fermions and bosons in
equilibrium at temperature $T$. Particle multiplicities are
given by
\begin{equation}
N_i = \frac{g_iV}{(2\pi\hbar)^3} \int^{\infty}_0 \frac{4 \pi p^2dp}
{{\mathrm exp}[(E_i-B_i\mu_B-S_i\mu_S)/T]\pm 1}.
\label{eq1}
\end{equation}
Here $g_i$ is the spin-isospin degeneracy factor of particle $i$, $E_i$, $B_i$ 
and
$S_i$ are the single particle energy, baryon number and strangeness, and 
$\mu_B$ and
$\mu_S$ are the baryon and strangeness chemical potentials (the electric
chemical potential has been neglected). $V$ is the volume
of the box. For resonances with a finite width, Eq.(\ref{eq1})
is folded by an integration over the mass distribution $\rho_i(m)$:
\begin{equation}
N_i = \frac{g_iV}{(2\pi\hbar)^3} \int \rho_i(m) dm \int^{\infty}_0 
\frac{4 \pi p^2dp}{{\mathrm exp}[(E_i(m)-B_i\mu_B-S_i\mu_S)/T]\pm 1}~~,
\label{eq2}
\end{equation}
For $\rho_i(m)$, the Breit-Wigner mass distribution of resonance $i$ is used. 
All hadron species used in the UrQMD model \cite{urqmd} have 
been included in the present statistical model. The UrQMD model uses a
stochastic collision term, and soft or hard core effects are neglected  
for all 
different particles (cascade calculations). Hence, excluded volume (Van der
Waals) correction \cite{goren} has not been included into 
Eqs.(\ref{eq1},\ref{eq2}). The excluded volume decreases the energy density and
changes the EoS as well as transport coefficients considerably in the high
density limit. The inclusion into a relativistic transport theory in
nontrivial, the non-relativistic Chapmann-Enskog expansion cannot simply be
extended due to causality problems arising in the relativistic case.

The strategy
adopted to determine the three parameters of the statistical model ($T$, 
$\mu_B$,
and $\mu_S$) is the following: Instead of fitting these parameters to some
particular particle multiplicities and energy spectra \cite{stat,stat1,goren},
the three parameters of the statistical model 
are defined by the input used for the corresponding UrQMD box calculations. 
For fixed energy density,
baryon density and strangeness density, the parameters are defined such that:

\begin{eqnarray}
\epsilon &=& \sum_{i} \frac{g_i}{(2\pi\hbar)^3}  
\int^{\infty}_0 \frac{E_i(m)4 \pi  p^2dp}
{{\mathrm exp}[(E_i(m)-B_i\mu_B-S_i\mu_S)/T]\pm 1} \label{eq3}~~;\\
\rho_B &=& \sum_{i} B_i \frac{N_i}{V} \label{eq4}~~;\\
\rho_S &=& \sum_{i} S_i \frac{N_i}{V}~~,
\label{eq5}
\end{eqnarray}
Here $i$ runs over all hadron species (and their anti-particles).  
The temperature varies with the energy density obtained in this model 
(at $\rho_B =$ 0.16 fm$^{-3}$
and $\rho_S =$0 fm$^{-3}$) as shown by the solid curve in Fig. 3. 
The curve exhibits a continuous rise of the temperature with the energy
density, in good agreement with UrQMD box calculations without strings and
many-body decays (solid circles).

However, in the above
sums, string degrees of freedom (which can be considered as heavy mass
resonances with small life times) are not taken into account. Therefore, 
a direct
comparison with the UrQMD model (which includes these degrees of freedom) shall
yield different results at high energy densities. 
The same results from the UrQMD box model and the corresponding statistical
model can only be expected if strings (or higher mass resonances) are included 
in the statistical model. 
For this,
the Hagedorn
mass spectrum \cite{hage} for the strings given by:
\begin{equation}
\rho^{(s)}(m) = \rho^{(s)}_0m^{a_H}{\mathrm exp}(m/T_H)
\label{hagedorn}
\end{equation}
must be included. We fix the Hagedorn temperature $T_H$ to the limiting 
temperature
obtained in standard UrQMD box calculations, $T_H = 130$ MeV. Moreover, the
constants $\rho^{(s)}_0$ and $a_H$ have been fixed such that to obtain the same
equation of state $\epsilon(T)$ as in the UrQMD box model (open diamonds in
Fig. 3). Their values are: $\rho^{(s)}_0 =$ 500 MeV$^2$ and $a_H = -3$. 

Six different
kinds of strings are excited within the UrQMD model, namely baryon strings with
different strangeness content ($B=1$ and $S=0,-1,-2,-3$) and meson strings
with strangeness 0 
or 1 ($B=0$ and $S=0,1$) and their
respective anti-strings (anti-baryon strings, $B=-1$ and $S=0,1,2,3$ and
anti-meson string, $B=0$ and $S=-1$). String multiplicities are then given by:
\begin{equation}
N^{(s)}_j = \frac{V}{(2\pi\hbar)^3} \int_{m_{min}}^{\infty} \rho^{(s)}(m) dm 
\int^{\infty}_0 \frac{4 \pi p^2dp}
{{\mathrm exp}[(E_j(m)-B_j\mu_B-S_j\mu_S)/T]\pm 1}~~,
\end{equation}
where the index $j$, $j = 1...6$, distinguishes the six kinds of string 
defined above 
(plus their respective
anti-strings). The lowest mass limit in the mass integration is fixed to
$m_{min} = 1.7$ GeV, the lowest mass of the strings excited in UrQMD 
calculations. The upper limit is fixed to $m_{max} = 1000$ GeV.
The variation of the temperature with the energy density in the
statistical model of hadrons and strings (discrete and continuous mass
spectrum, respectively) is shown in Fig. 3 by the dashed line. 

Figure 4 shows particle multiplicities of all baryon (top
panel) and meson (lower panel) species as obtained in UrQMD box calculation at 
$\epsilon = 1$ GeV/fm$^3$, $\rho_B = 0.16$ fm$^{-3}$ 
and $\rho_S = 0$ fm$^{-3}$ (club symbols). 
Also shown are hadron multiplicities resulting from the
statistical model where the continuum mass spectrum has been taken into account
according to Eq. (6) (open circles). 

From Eqs. (3-5), with the sums extended to include 
string degrees of freedom, we get from the statistical model the following
values for the intensive thermodynamical variables: 
$T = 129.98$ MeV, $\mu_B = 505$ MeV and $\mu_S = 130$ MeV.
Large discrepancies are observable between the UrQMD box model and the 
statistical
model for almost all hadron species. The statistical model overestimates almost
all baryon multiplicities (e.g. by a factor 5 for nucleons), whereas it 
underestimates drastically meson multiplicities (almost by an order of magnitude
for pions).

It appears that the difference between
the two models is due to the principle of 
detailed balance, which is violated when strings and many-body decays of
resonances (more than two outgoing particles)
are included into the UrQMD box calculations. It is obvious that detailed 
balance (which is assumed in the statistical model) must be
violated in many-body decays, if for the back reaction 
only binary collisions are considered, as in the
microscopic UrQMD simulation, which employs a Boltzmann limit collision kernel
(with molecular chaos). Chapmann-Enskog type extensions would exaggerate this
effect even more. String
and many-body decays lead to an enhancement of light mesons, mostly pions,
resulting in non-zero effective chemical potentials. 

Figure 5 shows 
hadron abundances as obtained in UrQMD box simulations (club
symbols) if
string degrees of freedom are excluded. At the same time, also all other 
many-body ($n>2$) decays are suppressed. The
results of the corresponding statistical model are shown in the same figure by
open circles for $T = 184$ MeV, $\mu_B = 67$ MeV and 
$\mu_S = 48$ MeV from Eqs. (3-5). The sums in the equations run only over 
hadron degrees of
freedom without the continuous mass spectrum. 
Good agreement 
between the two models is observed, although there are still some differences 
for the heaviest resonances. The particles with
the highest production probabilities (nucleons, deltas, pions, kaons, etc...) 
however, agree much better than in Fig. 4. Note that $\eta$'-
and f$_1$-mesons are missing in Fig. 5. This is because these particles can be 
produced in
UrQMD simulations only via string decays, which have now been explicitly
canceled.

Energy spectra of the particles shown in Fig. 6 are also nicely 
reproduced a temperature of 180 MeV. This $T$-value is 
in agreement with the temperature obtained
in the statistical model for the same input $\epsilon$, $\rho_B$ and $\rho_S$
(see above). 
The dependence of the temperature on the energy density
as obtained in the UrQMD simulations, but without strings and many-body 
decays, is
shown in Fig. 3 by the solid circles. The UrQMD model and the statistical model
of ideal
hadrons seem to agree when strings and many-body decays are excluded
in the UrQMD model. The inclusion of strings and many-body
decays of resonances results in a Hagedorn-like equation of state.

At this point, a comment is due. It is evident for both models (UrQMD and the
statistical model) that string degrees of freedom 
(a continuously increasing mass spectrum in the statistical model) become 
increasingly
relevant with increasing energy density. As an example, at
$\epsilon=1$ GeV/fm$^3$ and $\rho=0.16$ fm$^{-3}$, 
both models show that an important fraction of the energy density and of the
net baryon density is stored in string 
degrees of freedom. These string degrees of freedom must therefore be taken 
into account for interpreting heavy-ion experimental data. To date, this has
been totally ignored. 

In conclusion, we have studied equilibrium properties of relativistic 
hadronic matter in the
framework of the UrQMD model. The system is confined to a box with periodic
boundary conditions at fixed baryon density and zero net strangeness. 
Starting from
random initial conditions, particle multiplicities saturate after some time,
indicating chemical equilibration. Strange hadrons show much longer 
equilibration times than non-strange 
hadrons 
at low energy densities. This difference, however, disappears 
at energy densities $\epsilon > 500$ MeV/fm$^3$. 
The slopes of
all hadron spectra can be reproduced by Boltzmann fits with two
temperatures $T_B$ and $T_M$, indicating a complete thermalization of the 
system with two-body decay contributions. These
equilibrium properties allow us to study local thermodynamical equilibrium in
realistic heavy-ion collisions \cite{bravina1}. 

The  
equation of state of the hot, dense hadron medium (energy density versus 
temperature) is extracted from the UrQMD model.
The EoS appears to be Hagedorn-like, with a limiting temperature of about 
$T_H = $130$\pm$10 
MeV. When comparing the UrQMD results with those resulting from a statistical
model for ideal 
hadrons based on the grand
canonical ensemble, the two models agree much better (for hadron multiplicities 
and energy spectra) when strings and other
many-body decays are suppressed in the UrQMD model. 
The differences when strings and many-body decays are included, are quite 
pronounced ($>$50\%) at energy densities $\epsilon >$ 500 MeV/fm$^3$.
This is due to the Boltzmann collision kernel with its 
restriction to binary collisions in the UrQMD model. It results in a 
violation of the principle of detailed balance, which is the basis of 
the statistical
model when strings and many-body 
decays are taken into account. On the
other hand, strings are needed for a description of relativistic
heavy-ion data in terms of subsequent hadron-hadron collisions (and even for
simple first collision models) \cite{urqmd}. The strings carry a 
substantial fraction of the energy
density and the baryon density.Therefore, they should be taken into account 
in thermal models for comparisons to experimental data in 
relativistic heavy-ion collisions.


\begin{figure}
\centerline{\epsfysize=18cm \epsfbox{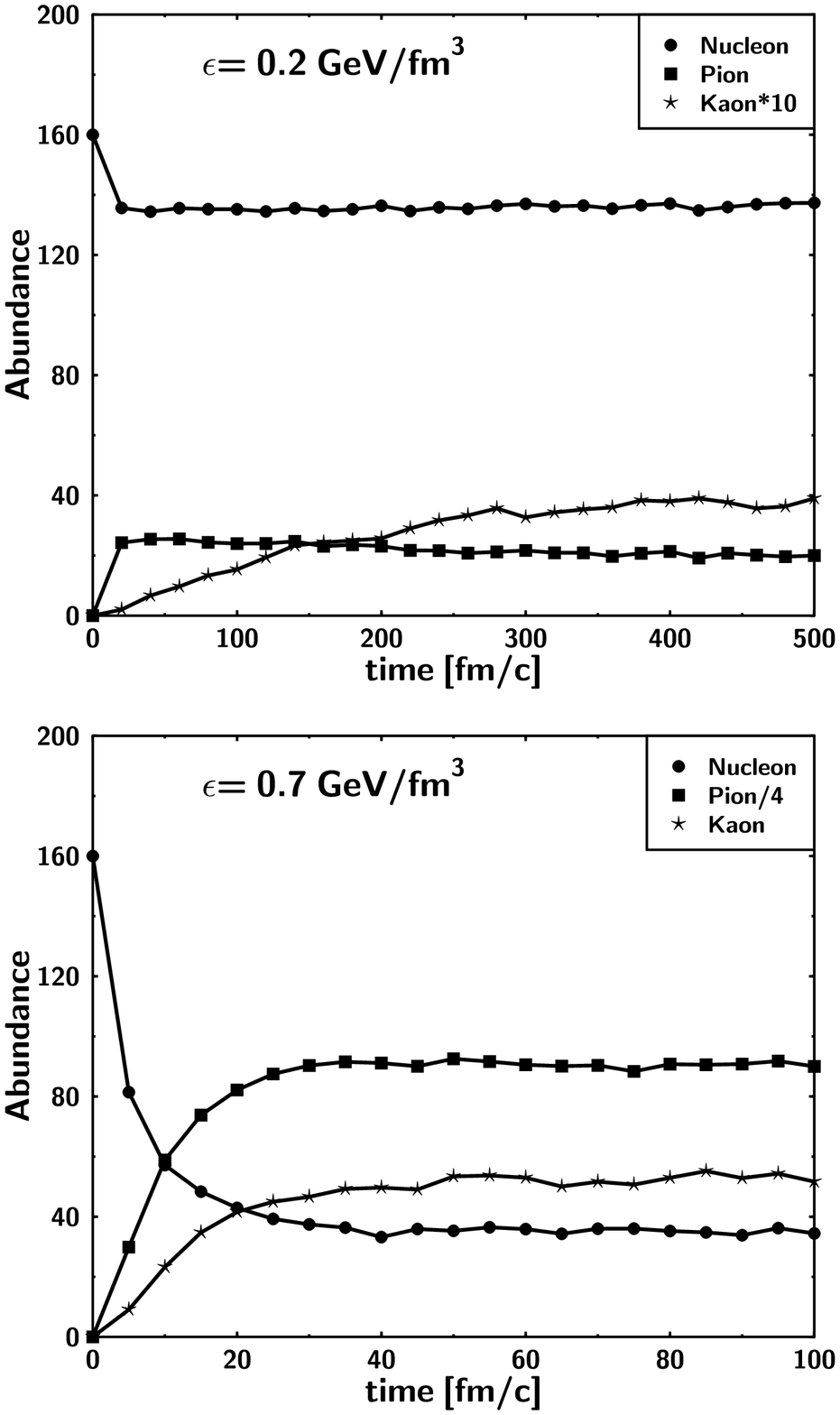}}
\caption{Time evolution of particle multiplicities for two different energy 
densities. The calculations are done at $\rho_B = 0.16$ fm$^{-3}$ and 
$\rho_S = 0$ fm$^{-3}$.}
\end{figure}

\begin{figure}
\centerline{\epsfysize=18cm \epsfbox{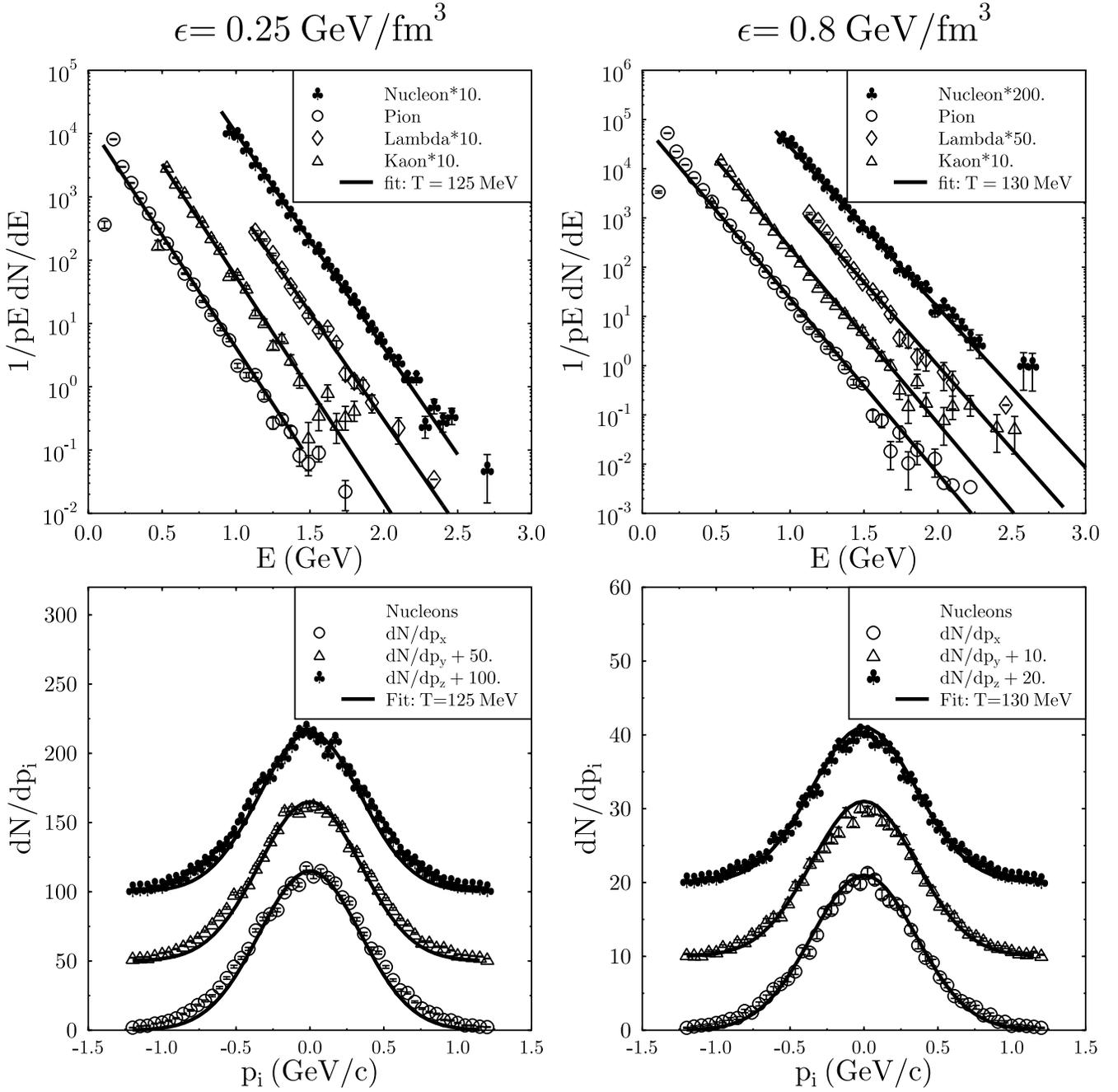}}
\caption{Energy spectra (top panels) of nucleons, pions, lambdas and kaons, 
and momentum distributions (low panels) of nucleons at two different energy
densities, 0.25 GeV/fm$^3$ (left part) and 0.8 GeV/fm$^3$ (right part).
The calculations are done at $\rho_B = 0.16$ fm$^{-3}$ and 
$\rho_S = 0$ fm$^{-3}$.}
\end{figure}

\begin{figure}
\centerline{\epsfysize=10cm \epsfbox{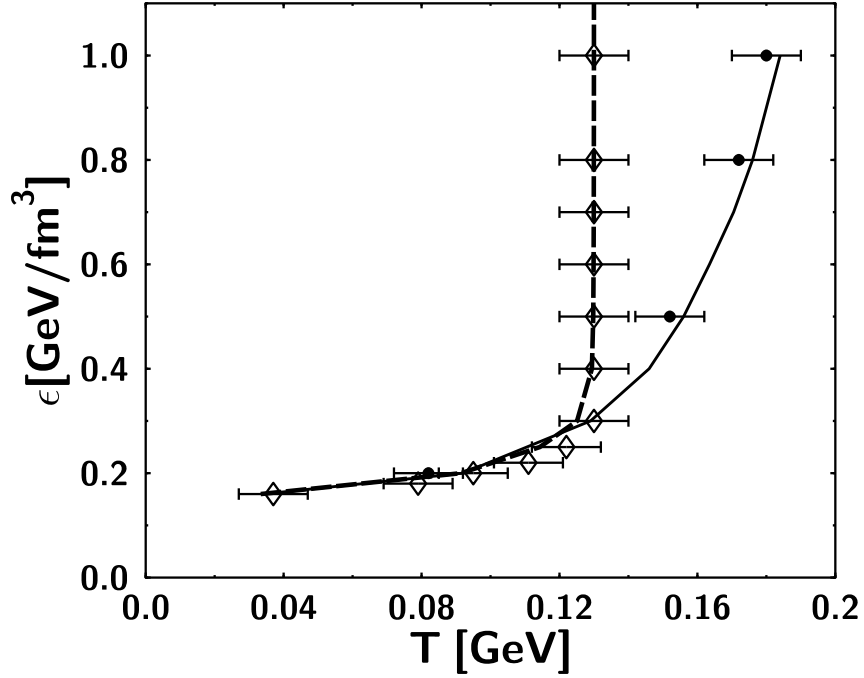}}
\caption{Equation of State. The energy density $\epsilon$ is plotted 
versus the temperature $T$. The symbols show
the results of UrQMD box calculations with (open diamonds) and without (solid
circles) strings and many-body ($n>$2) decays. The lines show the EoS of a
quantum statistical model of ideal hadrons (solid line) plus continuous mass
spectrum (dashed line) (see text).
The calculations are done at $\rho_B = 0.16$ fm$^{-3}$ and 
$\rho_S = 0$ fm$^{-3}$.}
\end{figure}

\begin{figure}
\centerline{\epsfysize=16cm \epsfbox{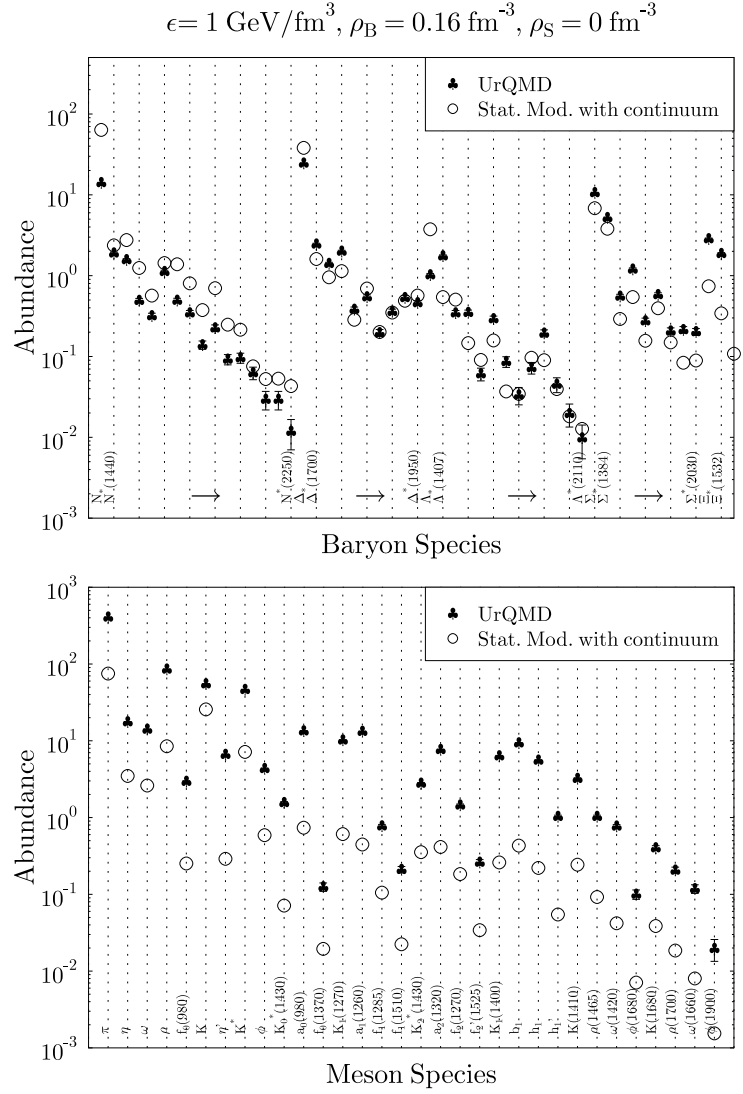}}
\caption{Absolute particle abundances as obtained in UrQMD box calculations 
(club symbols) and
in the statistical model (open circles) where high mass resonances from the
continuum have been taken into account.  }
\end{figure}

\begin{figure}
\centerline{\epsfysize=16cm \epsfbox{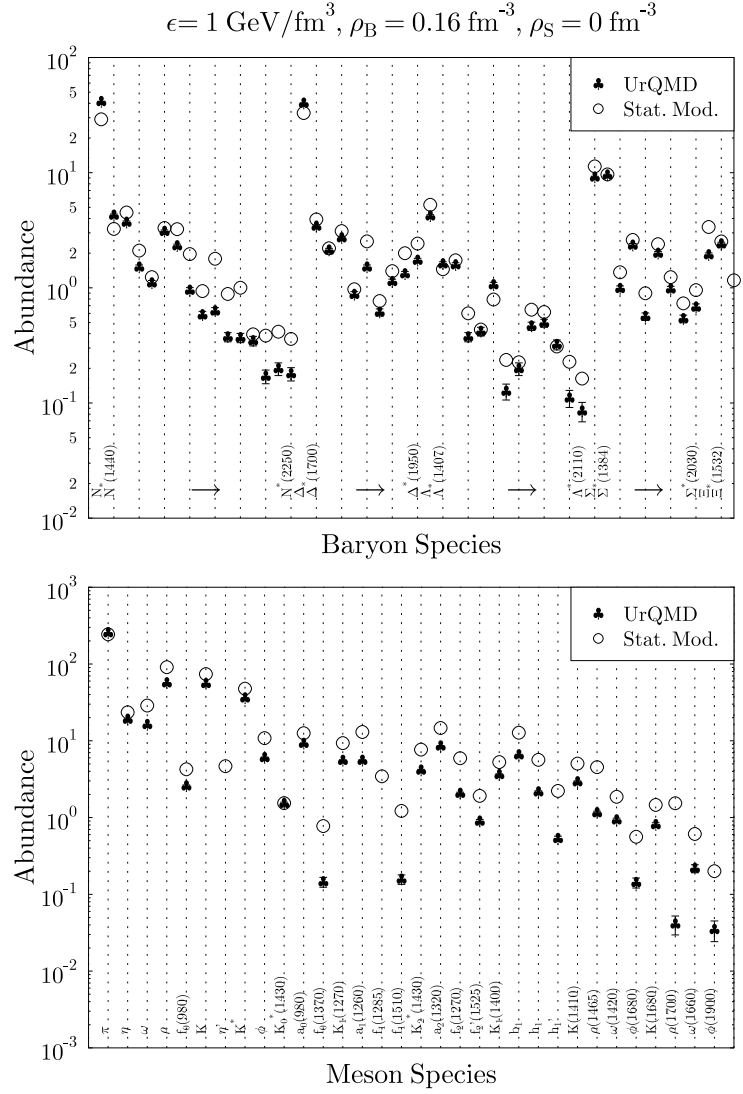}}
\caption{Absolute particle abundances from UrQMD box calculations 
(club symbols) and from the 
statistical model (open circles). In UrQMD calculations, strings and many-body 
decays have been suppressed while no continuum has been
taken into account in the statistical model. }
\end{figure}

\begin{figure}
\centerline{\epsfysize=11cm \epsfbox{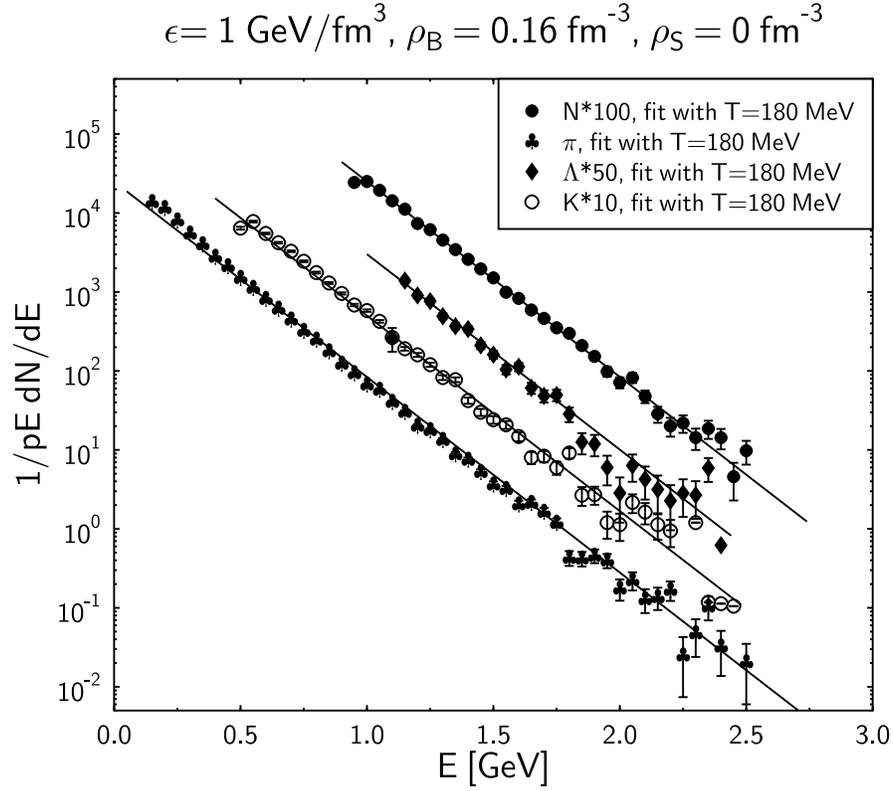}}
\caption{Energy spectra for nucleons, pions, lambdas and kaons resulting from
UrQMD box calculations without strings and many-body decays. 
All spectra are fitted
by Boltzmann distributions with a single temperature $T = 180$ MeV.}
\end{figure}

\end{document}